\documentclass[12pt]{article}
\usepackage{graphicx}
\usepackage{amsmath}
\usepackage{amssymb}
\usepackage{cite}
   
\newcommand\Ref[1] {Ref.\,\cite{#1}}
\newcommand\Refs[1] {Refs.\,\cite{#1}}
\newcommand\eqn[1] {Eq.\,(\ref{#1})}

\newcommand\fig[1] {Figure~\,{\ref{#1}}}

\newcommand\tab[1] {Table~\ref{#1}}
\newcommand\subtitle[1] {\noindent{\bf #1}}

\def\beq{\begin{equation}}
\def\eeq{\end{equation}}
\def\beqq{\begin{equation*}}
\def\eeqq{\end{equation*}}
\def\bsp#1\esp{\begin{split}#1\end{split}}
\def\bal#1\eal{\begin{align}#1\end{align}}

\newcommand\rd   {\mathrm{d}}
\newcommand\aS   {\alpha_{\mathrm{S}}}

\newcommand{\CF} {C_{\mathrm{F}}}
\newcommand{\CA} {C_{\mathrm{A}}}
\newcommand{\TR} {T_{\mathrm{R}}}

\newcommand{\nf} {n_{\mathrm{f}}}

\newcommand\zc   {z_{\mathrm{c}}}
\newcommand\cd   {\!\cdot\!}

\newcommand\CNNLO {CoLoRFulNNLO\,}

\newcommand\epem {electron-positron }

\begin{document}


\begin{titlepage}
\renewcommand{\thefootnote}{\arabic{footnote}}
\par \vspace{2cm}

\begin{center}
{\Large \bf Soft-drop event shapes in electron-positron \\[0.5em]
annihilation at next-to-next-to-leading order accuracy}
\end{center}
\par \vspace{10mm}
\begin{center}
{\bf Adam Kardos}$^{(a)}$,
{\bf G\'abor Somogyi}$^{(b)}$ 
and 
{\bf Zolt\'an Tr\'ocs\'anyi}$^{(a,b)}$\\

\vspace{5mm}

$^{(a)}$ University of Debrecen, Faculty of Science and Technology, Institute of Physics\\
$^{(b)}$ MTA-DE Particle Physics Research Group\\
\vspace{5mm} 
H-4010 Debrecen, PO Box 105, Hungary
\end{center}

\par \vspace{2cm}
\begin{center} {\large \bf Abstract} \end{center}
\begin{quote}
\pretolerance 10000
We present predictions for soft-drop event shapes of hadronic final states 
in electron-positron annihilation at next-to-next-to-leading order accuracy 
in perturbation theory obtained using the CoLoRFulNNLO subtraction method. 
We study the impact of the soft drop on the convergence of the perturbative 
expansion for the distributions of three event shape variables, the soft-drop 
thrust, the hemisphere jet and narrow jet invariant masses. We find that 
grooming generally improves perturbative convergence for these event 
shapes. This better perturbative stability, in conjunction with a reduced 
sensitivity to non-perturbative hadronization corrections makes soft-drop 
event shapes promising observables for the precise determination of the 
strong coupling at lepton colliders.
\end{quote}

\vspace*{\fill}
\begin{flushleft}
July 2018
\end{flushleft}
\end{titlepage}
\clearpage

\renewcommand{\thefootnote}{\fnsymbol{footnote}}


The precise determination of the strong coupling $\aS$ is important 
for improving our understanding of the fundamental interactions. For
instance, the value of $\aS$ has the largest effect on the running of
the effective potential of the Higgs field among the gauge couplings of
the standard model \cite{Degrassi:2012ry}. In principle, \epem colliders 
provide the cleanest laboratories to carry out such a measurement 
because all strongly interacting particles emerge only in the final state. 
At the LEP collider numerous observables were studied extensively
to determine the value of $\aS$ at various center-of-mass energies.  
A large class of such observables are distributions of event shape
variables such as thrust, $T$ \cite{Brandt:1964sa,Farhi:1977sg}.

Thrust is among the best studied event shapes and provides a good 
example for understanding the theoretical issues that need to be 
overcome for a precise determination of the strong coupling. The first 
of these relates to the precise perturbative description of the observable. 
The thrust distribution is known to NNLO precision in fixed-order perturbation 
theory \cite{GehrmannDeRidder:2007hr,Weinzierl:2009ms,DelDuca:2016csb,
DelDuca:2016ily}, while the resummation of large logarithms for small 
$\tau = 1-T$ has been carried out to N$^3$LL accuracy using SCET in 
\Ref{Becher:2008cf} where matched predictions at NNLO+N$^3$LL 
accuracy were also presented. Yet the perturbative corrections are not 
particularly small even at NNLO and one observes a significant difference 
between the predictions and measured data, especially in the peak 
region around $\tau \simeq 2.5\cdot 10^{-2}$ where the statistics of 
data are the best.

A second issue concerns the estimation of hadronization corrections 
that mostly account for the difference between the perturbative predictions 
and data. These corrections must either be extracted from data by comparison 
to Monte Carlo predictions or computed using analytic models and the lack of 
reliable predictions for hadronization from first principles hampers the precise 
determination of the strong coupling in this potentially clean environment. One 
possible way to improve on this situation is to look for observables for which
the hadronization corrections are much reduced as compared to traditional 
ones.  

Generic classes of observables with reduced non-perturbative corrections 
can be obtained by various methods of grooming \cite{Butterworth:2008iy,
Krohn:2009th,Ellis:2009su,Ellis:2009me,Dasgupta:2013ihk,Larkoski:2014wba}, 
which were originally developed for hadron colliders. However, understanding 
the structure of theoretical predictions for groomed jets is usually more 
complicated than for un-groomed ones. Nevertheless, for a particular type 
of grooming called soft drop \cite{Larkoski:2014wba}, significant progress 
to perform all-order calculations has been made \cite{Frye:2016aiz,Frye:2016okc,
Marzani:2017mva,Marzani:2017kqd,Kang:2018jwa}, although the resummation 
program at high perturbative orders still poses computational challenges.  
A recent development related to soft-drop jet observables is the definition 
of soft-drop thrust and related quantities in \epem annihilation \cite{Baron:2018nfz}, 
whose hadronization corrections are indeed significantly reduced by 
the soft drop.

In this letter we present fixed-order predictions at NNLO accuracy for 
three soft-drop groomed jet observables, the thrust, the hemisphere 
jet mass and the narrow jet mass measured in \epem  annihilation. 
We find that soft-drop grooming, in addition to reducing hadronization 
corrections, also leads to a better perturbative convergence of these 
quantities, making them promising observables for the precise 
determination of $\aS$.

The soft drop grooming technique was introduced in \Ref{Larkoski:2014wba} 
and defined for jets produced in lepton collisions in \Ref{Frye:2016aiz}.  
For the particular prescription that we employ here we refer to 
\Ref{Baron:2018nfz} where the definitions of the event shapes that we 
discuss -- (i) soft-drop thrust (version $T'_{\rm SD}$ that is free of the 
transition point in the soft-collinear region), (ii) hemisphere jet mass 
($e^{(2)}_2$) and (iii) narrow jet mass ($\rho$) -- are spelled out precisely. 
The soft-drop algorithm depends on two parameters $\zc$ and $\beta$. 
The effect of these parameters on hadronization corrections was studied 
in \Ref{Baron:2018nfz} where it was found that with increasing $\zc$ 
and decreasing $\beta$ (i.e., stronger grooming) the hadronization 
corrections to the soft-drop thrust are much reduced over a wide range 
of the event shape. But such changes in the grooming parameters 
also reduce the cross section. Thus, in addition to the small hadronization 
corrections, a further criterion for determining the optimal value of 
$\zc$ and $\beta$ is to avoid the loss of too much data.

The precision of $\aS$ determination is also influenced by the convergence 
of the perturbative series for the observable. Hence, it is important to 
examine how grooming affects the perturbative stability of the predictions. 
In order to assess this, we choose four pairs of $(\zc,\beta)$ values, 
$\{(0.05,1), (0.05,0), (0.1,1), (0.1,0)\}$, and compute the NLO and NNLO 
$K$-factors defined by ratios of distributions of the observable $O$ as
\beq
K_{\rm NLO}(\mu) =
\frac{\rd \sigma_{\rm NLO}(\mu)}{\rd O}
\bigg/
\frac{\rd \sigma_{\rm LO}(Q)}{\rd O}
\,,\quad
K_{\rm NNLO}(\mu) = 
\frac{\rd \sigma_{\rm NNLO}(\mu)}{\rd O}
\bigg/
\frac{\rd \sigma_{\rm LO}(Q)}{\rd O}
\,.
\label{eq:Kfactors}
\eeq
The normalization above is chosen such that the leading-order cross sections 
in the denominators are always computed at the default renormalization scale 
$\mu=Q$, independently of $\mu$. The closer the $K$-factors are to unity, the 
better the convergence of the perturbative series.\footnote{We have checked 
that the $K$-factors depend on the grooming parameters smoothly within the 
range $(\zc,\beta) \in [0.05,0.1]\times [0,1]$.}

The perturbative expansion of the differential distribution of an event shape $O$ 
at some arbitrary renormalization scale $\mu$ can be written to NNLO accuracy 
as 
\beq
\frac{O}{\sigma_0} \frac{\rd \sigma(\mu)}{\rd O} =
\frac{\aS(\mu)}{2 \pi} A_O(\mu) + \bigg(\frac{\aS(\mu)}{2 \pi} \bigg)^2 B_O(\mu) 
+ \bigg(\frac{\aS(\mu)}{2 \pi} \bigg)^3 C_O(\mu)  + {\mathcal O}(\aS^4)
\label{eq:NNLO}
\eeq
where $\sigma_0$ is the leading-order perturbative prediction for the process 
$e^+e^- \to q\bar{q}$. The dependence of the expansion coefficients on the 
value of the observable $O$ is understood, but suppressed. In practice it suffices 
to compute the functions $A_O(\mu)$, $B_O(\mu)$ and $C_O(\mu)$ at one 
particular value of the renormalization scale, since scale dependence is easily 
restored using the renormalization group equation for the strong coupling. 
Choosing the center-of-mass energy $Q$ as the default renormalization scale 
and denoting the perturbative coefficients at this scale as $A_O$, $B_O$ and 
$C_O$, we find
\beq
\label{eq:coeff_scale}
\bsp
A_O(\mu) &= A_O\,,\qquad
B_O(\mu)  = B_O + A_O \beta_0 \ln \xi\,,\\
C_O(\mu) &= C_O + 2 B_O \beta_0 \ln \xi 
+ A_O \bigg(\frac{1}{2} \beta_1 \ln \xi + \beta_0^2 \ln^2 \xi \bigg) \,,
\esp
\eeq
with $\xi \equiv \mu/Q$. The first two coefficients of the QCD $\beta$ function 
are \cite{Tarasov:1980au}
\beq
\label{eq:betas}
\beta_0 = \frac{11 \CA - 4 \TR \nf}{3}
\qquad\mbox{and}\qquad
\beta_1 = \frac{34}{3} \CA^2 - \frac{20}{3} \CA \TR \nf - 4 \CF \TR \nf
\,.
\eeq
We note that the expansion coefficients at the default renormalization scale, 
$A_O$, $B_O$ and $C_O$, are independent of the collision energy and the 
distribution of the observable at $\mu=Q$ depends on $Q$ only through the 
strong coupling $\aS(Q)$. 

The perturbative coefficients were computed using the \CNNLO method 
that was also used to calculate a variety of three-jet event shapes in \epem 
annihilation at NNLO accuracy \cite{DelDuca:2016csb,DelDuca:2016ily,
Tulipant:2017ybb}. Details of the formalism can be found in 
\Refs{Somogyi:2006da,Somogyi:2006db,DelDuca:2016ily}.

%
%
\begin{figure}
\begin{center}
\includegraphics[width=0.49\textwidth]{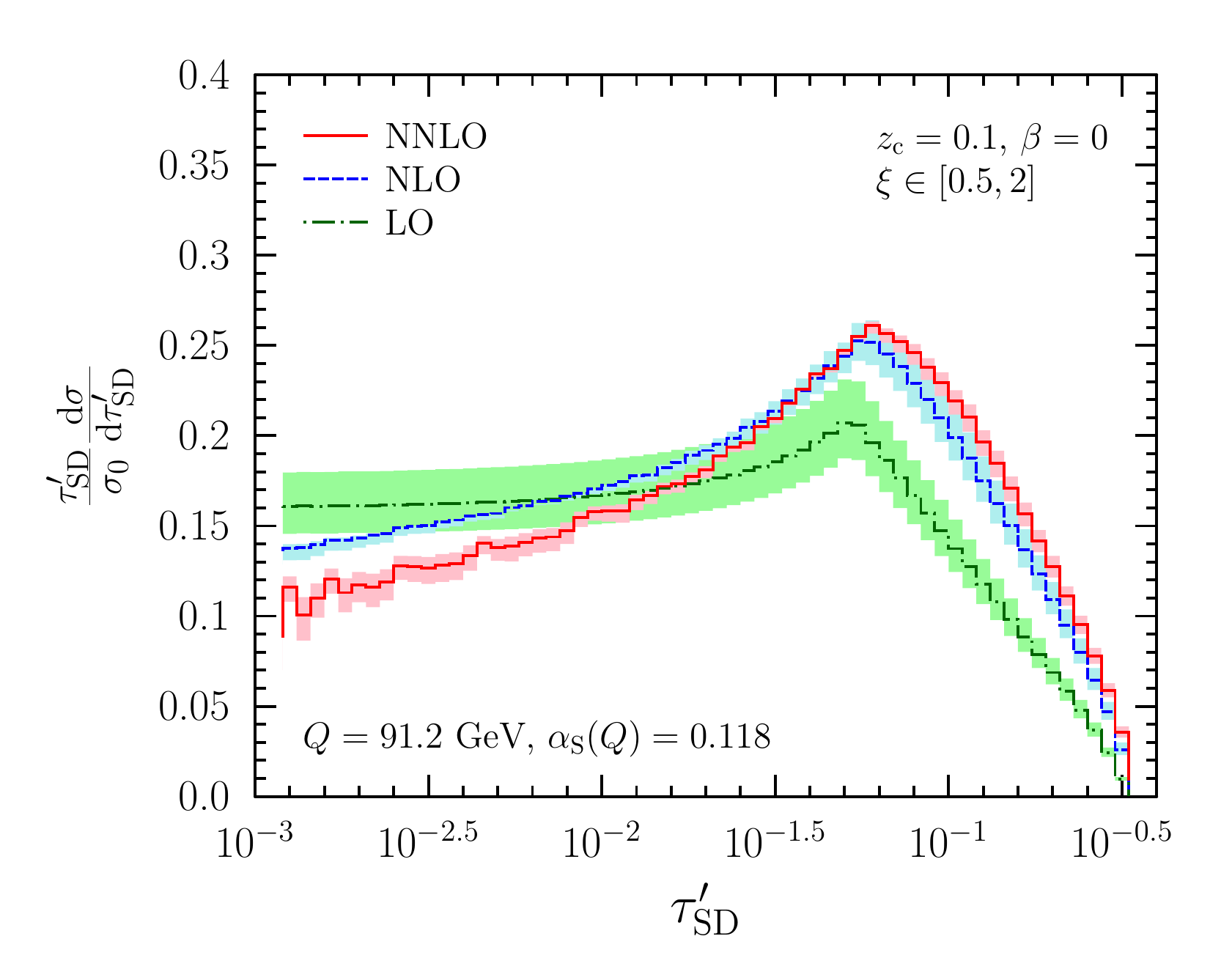}
\includegraphics[width=0.49\textwidth]{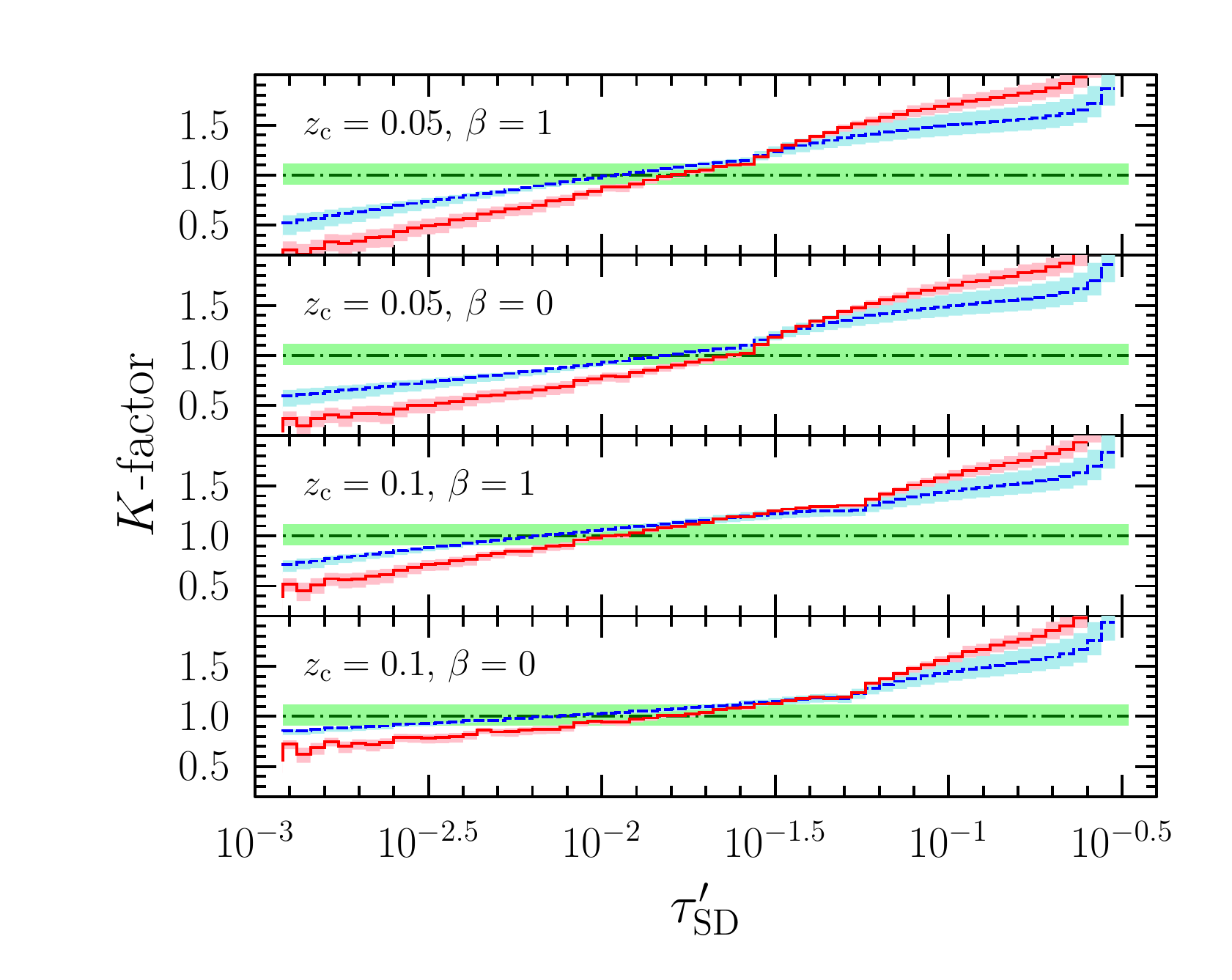}
\end{center}
\caption{\label{fig:tau} 
Predictions for the soft-drop thrust for grooming parameters $\zc=0.1$ and $\beta=0$ 
at LO (dot-dashed green), NLO (dashed blue) and NNLO (solid red) accuracy (left) 
and $K$-factors for various $\zc$ and $\beta$ values as indicated in the plots (left).
The bands represent renormalization scale variation in the range $\xi =\mu/Q \in [0.5, 2]$.}
\end{figure}

We begin the presentation of our results by considering the soft-drop thrust. 
Predictions for the distribution of $\tau'_{\mathrm{SD}} = 1 - T'_{\mathrm{SD}}$ 
for the center-of-mass energy of $Q=91.2$~GeV and grooming parameters 
$\zc=0.1$ and $\beta=0$ are shown on the left panel of \fig{fig:tau} at LO 
(dot-dashed green), NLO (dashed blue) and NNLO (solid red) accuracy. The 
value of the strong coupling was chosen as $\aS(M_Z) = 0.118$. The bands 
correspond to varying the renormalization scale in the range $\mu \in [Q/2,2Q]$. 
We also present the perturbative coefficients $A_{\tau'_{\mathrm{SD}}}$, 
$B_{\tau'_{\mathrm{SD}}}$ and $C_{\tau'_{\mathrm{SD}}}$ computed at 
$\mu=Q$ in \tab{tab:tau} for $(\zc,\beta) = (0.1,0)$, tabulated on a linear 
scale in $\tau'_{\mathrm{SD}}$. We observe the  very good numerical 
stability of our NNLO computation over the full range of the observable 
considered. We have checked that this stability does not depend on the 
values of $\zc$ and $\beta$.

Next, we investigate how the convergence of the perturbative prediction  
depends on the grooming parameters. On the right panel of \fig{fig:tau} 
we present the $K$-factors at NLO (dashed blue) and NNLO (solid red) 
as defined in \eqn{eq:Kfactors} for four pairs of $(\zc,\beta)$ values, 
$\{(0.05,1), (0.05,0), (0.1,1), (0.1,0)\}$. The constant LO $K$-factor 
(dot-dashed green) is also shown for visual reference.\footnote{Note 
however that the LO distribution itself depends significantly on the 
grooming parameters.} In general we find that milder grooming leads 
to larger change from order to order in perturbation theory. Thus, grooming 
improves perturbative convergence as one might expect. In the ranges 
considered, the dependence of the $K$-factors on $\beta$ is seen to 
be milder than their dependence on $\zc$. We observe that for 
$\tau'_{\mathrm{SD}} \gtrsim 10^{-2}$, i.e., in the range where the 
bulk of the data lies, the most stable perturbative prediction is obtained 
for $\zc=0.1$ and $\beta=0$.

%
%
\begin{figure}
\begin{center}
\includegraphics[width=0.49\textwidth]{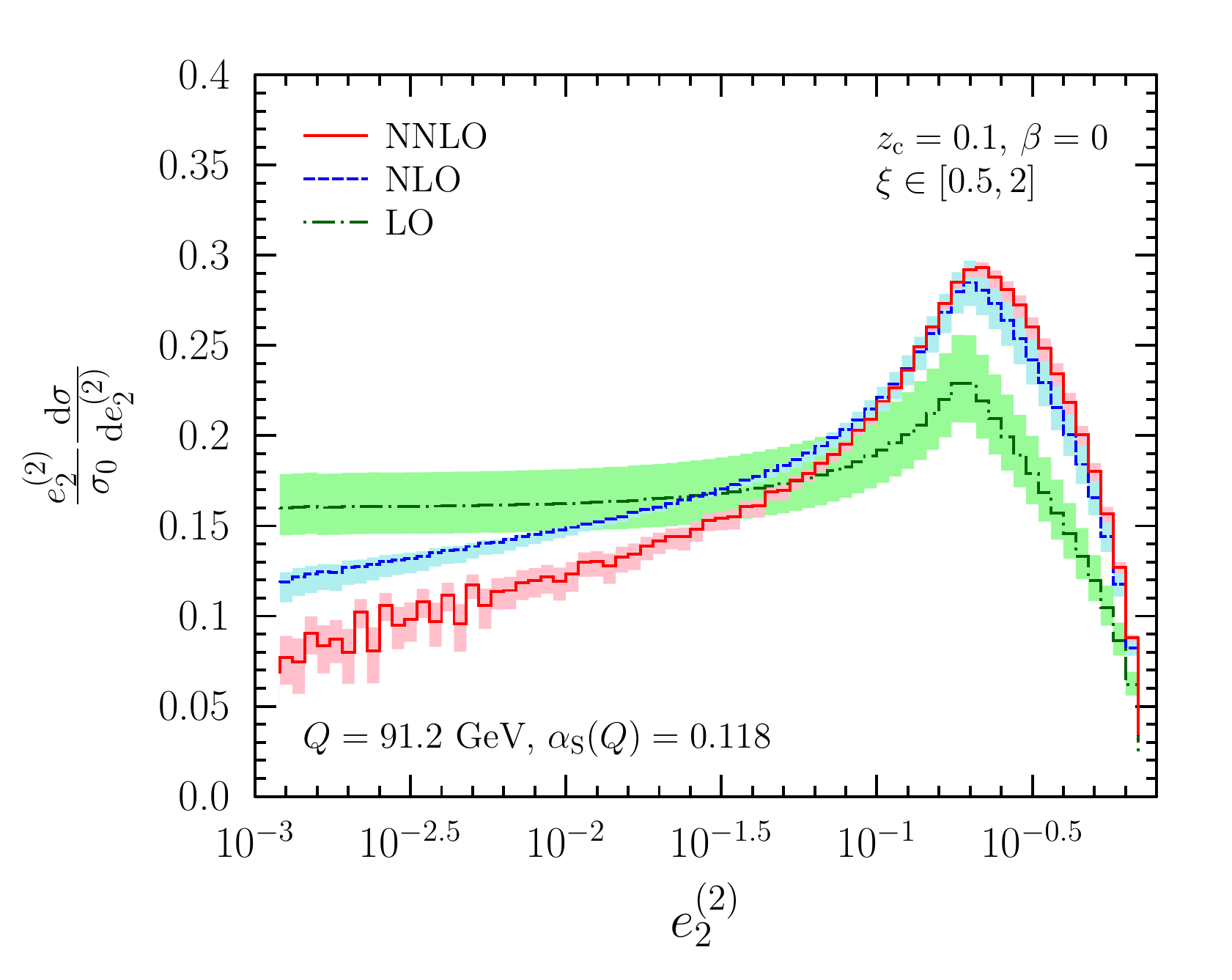}
\includegraphics[width=0.49\textwidth]{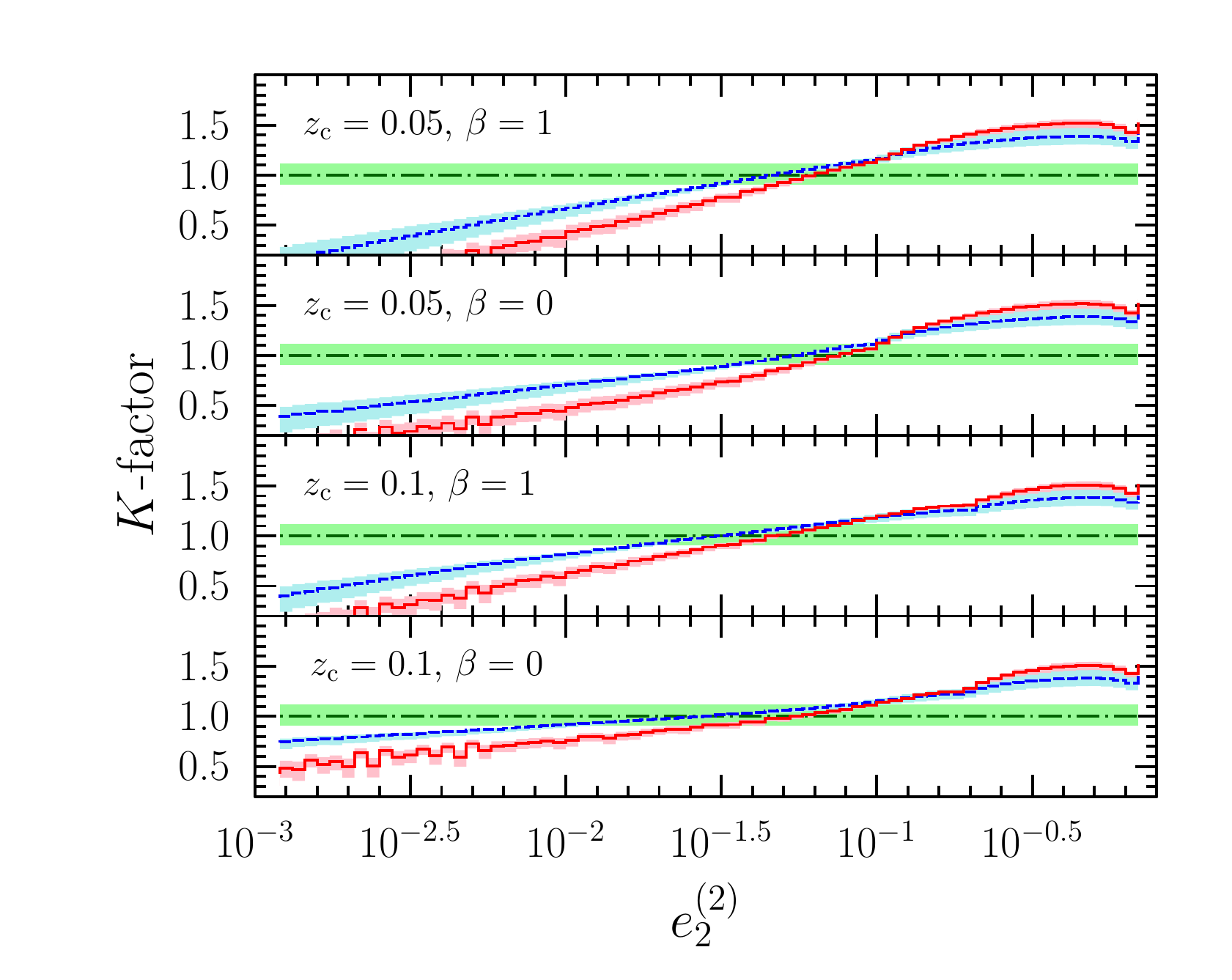}
\end{center}
\caption{\label{fig:e22} 
Predictions for the soft-drop hemisphere jet mass for grooming parameters 
$\zc=0.1$ and $\beta=0$ at LO (dot-dashed green), NLO (dashed blue) and 
NNLO (solid red) accuracy (left) and $K$-factors for various $\zc$ and $\beta$ 
values as indicated in the plots (left). The bands represent renormalization 
scale variation in the range $\xi =\mu/Q \in [0.5, 2]$.}
\end{figure}

Turning to the soft-drop hemisphere jet mass, in \fig{fig:e22} we present our 
perturbative predictions for the distribution of $e_2^{(2)}$ at LO, NLO and 
NNLO accuracy for $\zc = 0.1$ and $\beta=0$ on the left panel. The $K$-factors 
corresponding to the same set of $(\zc,\beta)$ values as for the soft-drop thrust 
are shown on the right panel. We also record in \tab{tab:e22} the values of the 
perturbative coefficients $A_{e_2^{(2)}}$, $B_{e_2^{(2)}}$ and $C_{e_2^{(2)}}$ 
computed at $\mu=Q$ for $(\zc,\beta) = (0.1,0)$, tabulated on a linear scale 
in $e_2^{(2)}$. We again observe the very good numerical stability of our 
NNLO predictions.

In general, we find that also for the soft-drop hemisphere jet mass, stronger 
grooming leads to an improved convergence of the perturbative predictions. 
In fact, the perturbative expansions of the distributions for soft-drop thrust 
and hemisphere jet mass behave very similarly with somewhat larger 
$K$-factors for the latter. We again find that choosing $\zc=0.1$ and 
$\beta=0$ leads to the perturbatively most stable predictions.

\begin{figure}
\begin{center}
\includegraphics[width=0.49\textwidth]{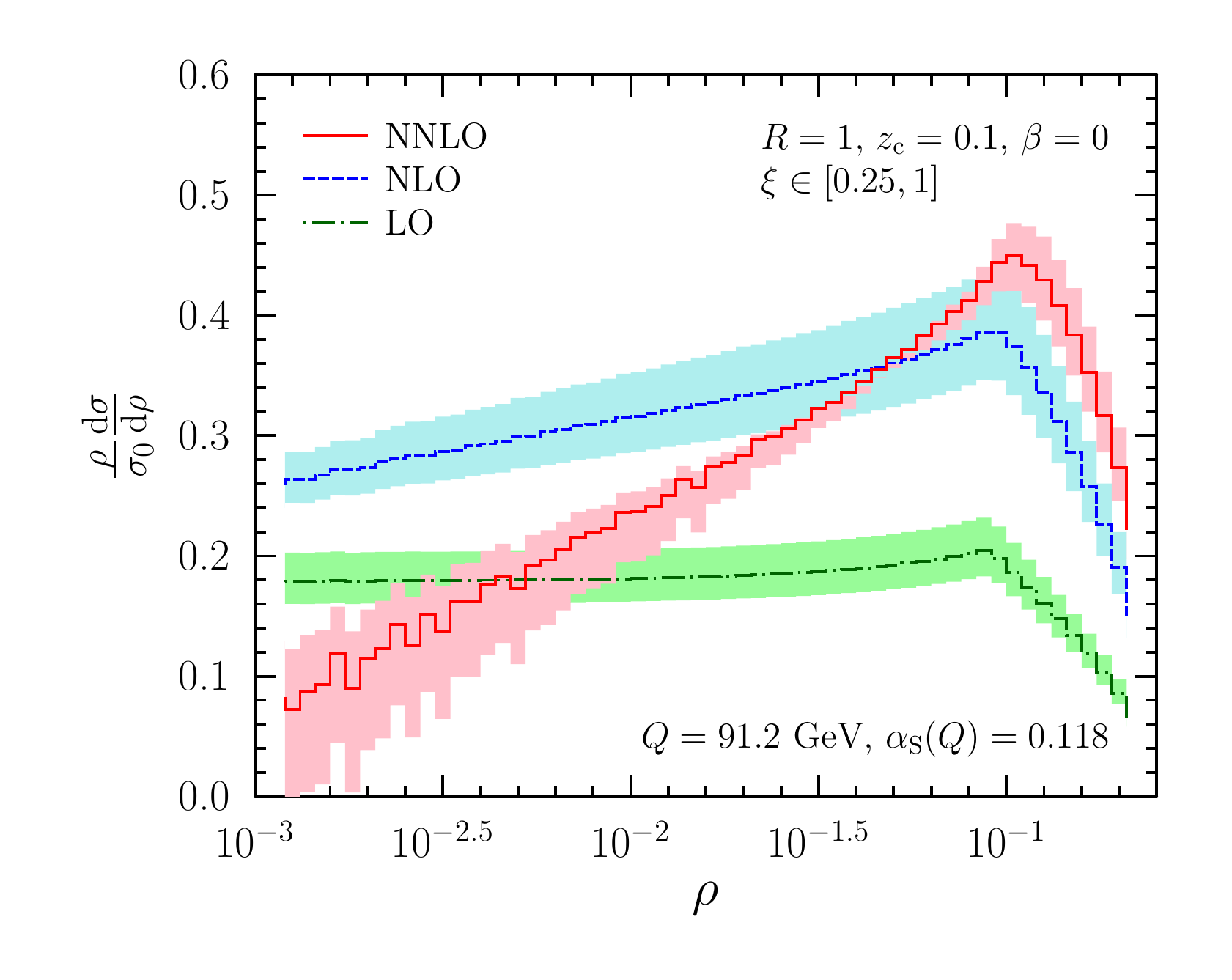}
\includegraphics[width=0.49\textwidth]{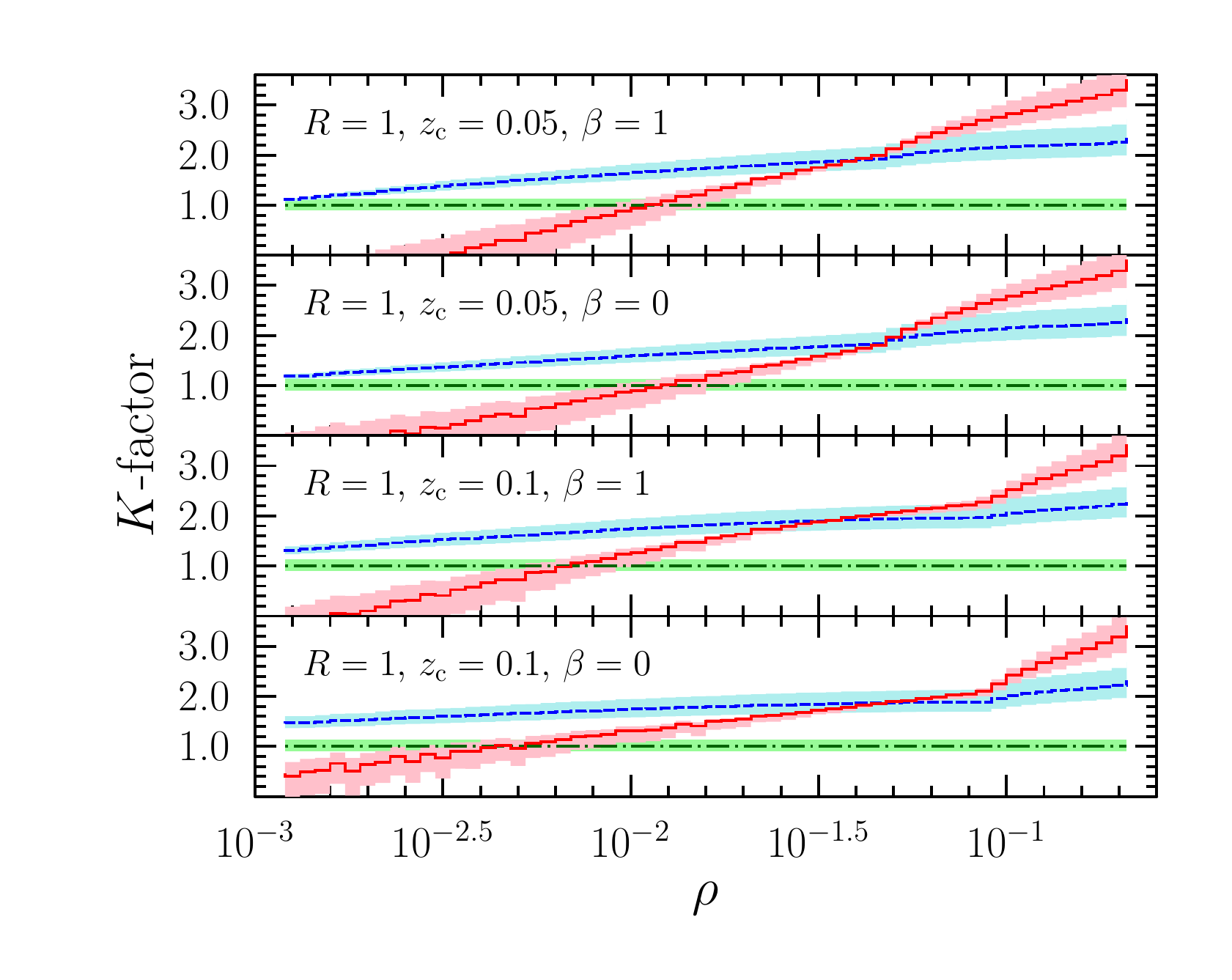}
\end{center}
\caption{\label{fig:rho} 
Predictions for the soft-drop narrow jet mass for jet radius $R=1$ and 
grooming parameters $\zc=0.1$ and $\beta=0$ at LO (dot-dashed green), 
NLO (dashed blue) and NNLO (solid red) accuracy (left) and $K$-factors 
for various $\zc$ and $\beta$ values as indicated in the plots (left). The bands 
represent renormalization scale variation in the range $\xi =\mu/Q \in [0.25, 1]$.}
\end{figure}

Last, we investigate the soft-drop narrow jet mass. The fixed-order predictions 
for the distribution of $\rho$ at LO, NLO and NNLO are shown on the left panel 
of \fig{fig:rho}, for jet radius $R=1$ (jets were defined using the anti-$k_t$ 
algorithm \cite{Cacciari:2008gp,Cacciari:2011ma}) and grooming parameters 
$\zc=0.1$ and $\beta=0$. In \Ref{Baron:2018nfz} it was shown that the natural 
hard scale for this observable is $\mu = \frac{Q R}{2}$, hence in \fig{fig:rho} 
we have set the central scale to $\mu = Q/2$. Nevertheless, in \tab{tab:rho} 
we present the expansion coefficients $A_{\rho}$, $B_{\rho}$ and $C_{\rho}$ 
computed at the default renormalization scale of $\mu=Q$ for $R=1$ and 
$(\zc,\beta) = (0.1,0)$, tabulated on a linear scale in $\rho$. The numerical 
convergence of the NNLO calculation is again found to be very good.

%
%

Examining the $K$-factors for the narrow jet mass, shown on the left panel 
of \fig{fig:rho}, alongside those for soft-drop thrust and hemisphere jet mass, 
a new feature emerges. Although stronger grooming does improve the perturbative 
convergence from NLO to NNLO (i.e., the ratio $K_{\rm NNLO}/K_{\rm NLO}$ is 
closer to unity), but the NLO $K$-factor is actually seen to increase with increasing 
$\zc$ and decreasing $\beta$ (i.e., more grooming). This is readily understood 
from the definition of $\rho$. Since a leading-order computation involves only 
three massless partons in the final state, in order to obtain a narrow jet mass 
that is non-zero at LO requires not only that the three partons be clustered into 
two jets, but also that the clustering passes the soft-drop condition. Otherwise, 
all soft-dropped jets are massless. Starting from NLO, the extra real radiation 
allows for configurations where three or more partons cluster to form a single 
jet. Such a jet can remain massive even after the soft drop. Hence one expects 
that the LO contribution is reduced more by stronger grooming than higher-order 
corrections, leading to an increased NLO $K$-factor at larger $\zc$ and smaller $\beta$.

We conclude that although in general grooming leads to better converging 
perturbative predictions, the sizes of the radiative corrections depend on the 
observable, being smaller for the more inclusive ones. In particular, for the 
narrow jet mass the NLO $K$-factor is rather large and positive (the $K$-factors 
are even larger for smaller jet radii), indicating that the NNLO computation 
represents the first reliable prediction for this quantity.

Finally, we offer a brief comment regarding all-order resummation for the 
observables considered here. Clearly, for small enough values of the 
event shapes the resummation of logarithmic contributions is mandatory. 
However, for soft-drop thrust and hemisphere jet mass specifically, the higher-order 
corrections remain moderate down to $\tau'_{\mathrm{SD}} \simeq 10^{-3}$ 
and $\rho \simeq 10^{-3}$ if $\zc=0.1$ and $\beta=0$. This suggests 
that the fixed-order NNLO predictions may potentially be reliable in the 
region with the bulk of the data (e.g., $\tau'_{\mathrm{SD}} \gtrsim 10^{-2}$). 
Hence, it might be possible that resummation only becomes essential for such 
a small range of the shape variable, that over this range the contribution to the 
cross section becomes more or less negligible. As the hadronization corrections 
are also small for $\zc=0.1$ and $\beta=0$ (under 10\% for $\tau'_{\mathrm{SD}} 
\gtrsim 10^{-2}$ \cite{Baron:2018nfz}), the increased perturbative stability makes 
the soft-drop thrust and hemisphere jet mass with such grooming parameters 
promising event shapes for a precise determination of $\aS$, perhaps even 
without resummation of large logarithms.

In this letter we presented predictions for soft-drop groomed event shapes 
of hadronic final states in \epem annihilation at NNLO accuracy in perturbation 
theory. Our predictions for the perturbative coefficients show very good 
numerical stability over the complete ranges of the observables considered. 
We have also studied the impact of grooming on the convergence of the 
perturbative expansions and presented NLO and NNLO $K$-factors for several 
values of the grooming parameters. We observed that in general, grooming 
improves the perturbative convergence of the predictions. This is more pronounced 
for the more inclusive quantities of soft-drop thrust and hemisphere jet mass, 
and less so for the narrow jet mass. The increased perturbative stability, 
along with reduced hadronization corrections makes the soft-drop thrust 
and hemisphere jet mass appealing candidates for a precise determination 
of the strong coupling at lepton colliders.

\subtitle{Acknowledgments}

We thank V.~Theevwes for sharing the data files of the figures in \Ref{Baron:2018nfz}.
A.K. acknowledges financial support from the Premium Postdoctoral Fellowship 
program of the Hungarian Academy of Sciences. This work was supported by 
grant K 125105 of the National Research, Development and Innovation Fund in Hungary.

%
%
\begin{table}             
\caption{Perturbative coefficients for the soft-drop thrust $\tau'_{\rm SD}$ with 
$(\zc,\beta) = (0.1,0)$ at $\mu=Q$. $C_{\tau'_{\mathrm{SD}}}$ is presented 
only up to the kinematical limit of $B_{\tau'_{\mathrm{SD}}}$.}
\label{tab:tau}
\footnotesize
\centering{
\begin{tabular}{cccc|cccc}
\hline 
  $\tau'_{\mathrm{SD}}$ & $A_{\tau'_{\mathrm{SD}}}\cd 10^{-3}$ &  $B_{\tau'_{\mathrm{SD}}}\cd 10^{-4}$ & $C_{\tau'_{\mathrm{SD}}}\cd 10^{-5}$ &
  $\tau'_{\mathrm{SD}}$ & $A_{\tau'_{\mathrm{SD}}}\cd 10^{-3}$ &  $B_{\tau'_{\mathrm{SD}}}\cd 10^{-4}$ & $C_{\tau'_{\mathrm{SD}}}\cd 10^{-5}$ \\
\hline                          
$0.005$ & $80.4(2)$ & $-2017(4)$ & $1787(4)$        &  $0.225$ & $0.013122(3)$ & $0.0447(3)$ & $0.113(5)$    \\ 
$0.015$ & $0.62894(7)$ & $0.202(2)$ & $-1.34(3)$    &  $0.235$ & $0.011427(3)$ & $0.0393(3)$ & $0.101(4)$    \\ 
$0.025$ & $0.38657(4)$ & $0.249(2)$ & $-0.48(3)$    &  $0.245$ & $0.009890(2)$ & $0.0363(3)$ & $0.097(4)$    \\ 
$0.035$ & $0.28926(4)$ & $0.251(2)$ & $-0.11(2)$    &  $0.255$ & $0.008492(2)$ & $0.0327(3)$ & $0.081(4)$    \\ 
$0.045$ & $0.23794(3)$ & $0.234(2)$ & $0.03(2)$     &  $0.265$ & $0.007196(2)$ & $0.0292(3)$ & $0.079(3)$    \\ 
$0.055$ & $0.19907(3)$ & $0.238(2)$ & $0.08(2)$     &  $0.275$ & $0.005991(2)$ & $0.0263(2)$ & $0.075(3)$    \\ 
$0.065$ & $0.15468(2)$ & $0.255(2)$ & $0.22(2)$     &  $0.285$ & $0.004862(2)$ & $0.0235(2)$ & $0.067(3)$    \\ 
$0.075$ & $0.12311(2)$ & $0.234(1)$ & $0.32(2)$   &  $0.295$ & $0.003791(2)$ & $0.0205(2)$ & $0.059(2)$    \\ 
$0.085$ & $0.10021(2)$ & $0.2091(9)$ & $0.32(2)$    &  $0.305$ & $0.0027649(9)$ & $0.0179(2)$ & $0.057(2)$   \\ 
$0.095$ & $0.08299(2)$ & $0.1877(9)$ & $0.30(2)$    &  $0.315$ & $0.0017736(8)$ & $0.01478(7)$ & $0.048(2)$  \\ 
$0.105$ & $0.069666(9)$ & $0.1662(8)$ & $0.31(2)$   &  $0.325$ & $801.8(5) \cd 10^{-6}$ & $0.01244(5)$ & $0.0425(7)$ \\ 
$0.115$ & $0.059103(8)$ & $0.1482(8)$ & $0.28(1)$ &  $0.335$ & $53.4(2) \cd 10^{-6}$ & $0.00773(2)$ & $0.0363(3)$ \\ 
$0.125$ & $0.050600(7)$ & $0.1304(7)$ & $0.26(1)$ &  $0.345$ & $0$ & $0.0025277(7)$ & $0.02025(5)$         \\ 
$0.135$ & $0.043611(6)$ & $0.1167(6)$ & $0.227(9)$  &  $0.355$ & $0$ & $0.0012796(3)$ & $0.01016(3)$         \\ 
$0.145$ & $0.037822(6)$ & $0.1043(6)$ & $0.228(8)$  &  $0.365$ & $0$ & $680.7(3) \cd 10^{-6}$ & $0.00557(2)$         \\ 
$0.155$ & $0.032941(5)$ & $0.0925(6)$ & $0.209(8)$  &  $0.375$ & $0$ & $356.9(2) \cd 10^{-6}$ & $0.00301(1)$       \\ 
$0.165$ & $0.028799(5)$ & $0.0839(5)$ & $0.181(7)$  &  $0.385$ & $0$ & $177.5(1) \cd 10^{-6}$ & $0.001576(6)$      \\ 
$0.175$ & $0.025239(4)$ & $0.0749(5)$ & $0.163(7)$  &  $0.395$ & $0$ & $79.04(7) \cd 10^{-6}$ & $767(4) \cd 10^{-6}$       \\ 
$0.185$ & $0.022159(4)$ & $0.0668(4)$ & $0.153(6)$  &  $0.405$ & $0$ & $28.19(4) \cd 10^{-6}$ & $328(2) \cd 10^{-6}$       \\ 
$0.195$ & $0.019465(4)$ & $0.0604(4)$ & $0.141(6)$  &  $0.415$ & $0$ & $5.21(2) \cd 10^{-6}$ & $102.3(7) \cd 10^{-6}$      \\ 
$0.205$ & $0.017095(3)$ & $0.0540(4)$ & $0.130(6)$  &  $0.425$ & $0$ & $2.6(2) \cd 10^{-9}$ & $6.6(2) \cd 10^{-6}$    \\
$0.215$ & $0.014994(3)$ & $0.0494(4)$ & $0.116(5)$  \\
\hline 
\end{tabular} 
}
\end{table}

%
%
\begin{table}             
\caption{Perturbative coefficients for the narrow jet mass $\rho$ with $R=1$ 
and $(\zc,\beta) = (0.1,0)$ at $\mu=Q$. $C_{\rho}$ is presented only up to the 
kinematical limit of $B_{\rho}$.}
\label{tab:rho}
\footnotesize
\centering{
\begin{tabular}{cccc|cccc}
\hline 
  $\rho$ & $A_{\rho}\cdot 10^{-3}$ &  $B_{\rho}\cdot 10^{-4}$ & $C_{\rho}\cdot 10^{-5}$ &
  $\rho$ & $A_{\rho}\cdot 10^{-3}$ &  $B_{\rho}\cdot 10^{-4}$ & $C_{\rho}\cdot 10^{-5}$ \\
\hline                        
$0.005$ & $174.8(5)$ & $-2492(5)$ & $2415(4)$       & $0.195$ & $0.022163(7)$ & $0.1384(2)$ & $0.613(5)$   \\ 
$0.015$ & $0.6028(2)$ & $2.555(2)$ & $-2.25(5)$     & $0.205$ & $0.018671(6)$ & $0.1194(2)$ & $0.550(4)$   \\ 
$0.025$ & $0.35799(7)$ & $1.601(2)$ & $0.02(4)$     & $0.215$ & $0.015404(5)$ & $0.1017(2)$ & $0.473(4)$   \\ 
$0.035$ & $0.25785(5)$ & $1.1847(9)$ & $0.62(3)$    & $0.225$ & $0.012191(5)$ & $0.0855(1)$ & $0.419(3)$ \\ 
$0.045$ & $0.20315(4)$ & $0.9489(7)$ & $0.86(3)$    & $0.235$ & $0.008854(4)$ & $0.07012(8)$ & $0.356(3)$  \\ 
$0.055$ & $0.16856(4)$ & $0.7945(7)$ & $0.99(2)$    & $0.245$ & $0.004583(3)$ & $0.05802(5)$ & $0.307(2)$  \\ 
$0.065$ & $0.14480(3)$ & $0.6863(6)$ & $1.03(2)$    & $0.255$ & $0$ & $0.033405(8)$ & $0.2442(7)$          \\ 
$0.075$ & $0.12755(3)$ & $0.6061(6)$ & $1.06(2)$    & $0.265$ & $0$ & $0.014268(4)$ & $0.1290(3)$          \\ 
$0.085$ & $0.11445(3)$ & $0.5445(5)$ & $1.05(2)$    & $0.275$ & $0$ & $0.007721(3)$ & $0.0734(2)$          \\ 
$0.095$ & $0.09988(3)$ & $0.5029(5)$ & $1.18(2)$    & $0.285$ & $0$ & $0.004371(2)$ & $0.0437(2)$          \\ 
$0.105$ & $0.08443(2)$ & $0.4521(4)$ & $1.25(2)$    & $0.295$ & $0$ & $0.002488(2)$ & $0.0257(1)$        \\ 
$0.115$ & $0.07200(2)$ & $0.3995(4)$ & $1.22(1)$  & $0.305$ & $0$ & $0.0013994(8)$ & $0.01518(7)$        \\ 
$0.125$ & $0.06184(2)$ & $0.3519(4)$ & $1.175(9)$   & $0.315$ & $0$ & $766.6(6) \cd 10^{-6}$ & $0.00873(5)$        \\ 
$0.135$ & $0.05340(2)$ & $0.3091(3)$ & $1.099(8)$   & $0.325$ & $0$ & $393.7(4) \cd 10^{-6}$ & $0.00489(3)$        \\ 
$0.145$ & $0.04624(2)$ & $0.2718(3)$ & $1.016(8)$   & $0.335$ & $0$ & $177.5(3) \cd 10^{-6}$ & $0.00251(2)$        \\ 
$0.155$ & $0.04012(1)$ & $0.2391(3)$ & $0.930(7)$ & $0.345$ & $0$ & $63.1(2) \cd 10^{-6}$ & $0.00111(1)$      \\ 
$0.165$ & $0.034803(9)$ & $0.2095(2)$ & $0.850(6)$  & $0.355$ & $0$ & $12.47(6) \cd 10^{-6}$ & $326(4) \cd 10^{-6}$      \\ 
$0.175$ & $0.030127(8)$ & $0.1834(2)$ & $0.775(6)$  & $0.365$ & $0$ & $26(2) \cd 10^{-9}$ & $16.3(6) \cd 10^{-6}$    \\
$0.185$ & $0.025951(8)$ & $0.1599(2)$ & $0.686(5)$ \\
\hline 
\end{tabular} 
}
\end{table}

%
%

\begin{table}             
\caption{Perturbative coefficients for the hemisphere jet mass $e_2^{(2)}$ with 
$(\zc,\beta) = (0.1,0)$ at $\mu=Q$. $C_{e_2^{(2)}}$ is presented only up to the 
kinematical limit of $B_{e_2^{(2)}}$.}
\label{tab:e22}
\footnotesize
\centering{
\begin{tabular}{cccc|cccc}
\hline 
  $e_2^{(2)}$ & $A_{e_2^{(2)}}\cd 10^{-3}$ & $B_{e_2^{(2)}}\cd 10^{-4}$ & $C_{e_2^{(2)}}\cd 10^{-5}$ &
  $e_2^{(2)}$ & $A_{e_2^{(2)}}\cd 10^{-3}$ & $B_{e_2^{(2)}}\cd 10^{-4}$ & $C_{e_2^{(2)}}\cd 10^{-5}$ \\
\hline 
$0.005$ & $79.2(2)$ & $-2016(4)$ & $1792(4)$          & $0.425$ & $0.017936(4)$ & $0.03617(4)$ & $0.064(3)$      \\ 
$0.015$ & $0.6040(2)$ & $-0.1789(8)$ & $-2.31(4)$     & $0.435$ & $0.017151(4)$ & $0.03477(4)$ & $0.059(3)$      \\ 
$0.025$ & $0.35912(7)$ & $-0.0177(5)$ & $-1.14(3)$    & $0.445$ & $0.016391(4)$ & $0.03330(4)$ & $0.056(3)$      \\ 
$0.035$ & $0.25908(5)$ & $0.0304(4)$ & $-0.70(2)$     & $0.455$ & $0.015674(4)$ & $0.03190(4)$ & $0.054(3)$      \\ 
$0.045$ & $0.20438(4)$ & $0.0507(4)$ & $-0.44(2)$     & $0.465$ & $0.014986(4)$ & $0.03060(4)$ & $0.054(3)$      \\ 
$0.055$ & $0.16982(4)$ & $0.0615(3)$ & $-0.35(2)$     & $0.475$ & $0.014334(3)$ & $0.02924(3)$ & $0.049(2)$      \\ 
$0.065$ & $0.14617(3)$ & $0.0677(3)$ & $-0.22(2)$     & $0.485$ & $0.013701(3)$ & $0.02801(3)$ & $0.048(2)$      \\ 
$0.075$ & $0.12900(3)$ & $0.0724(3)$ & $-0.18(2)$     & $0.495$ & $0.013093(3)$ & $0.02679(3)$ & $0.048(2)$      \\ 
$0.085$ & $0.11589(3)$ & $0.0750(2)$ & $-0.12(2)$     & $0.505$ & $0.012514(3)$ & $0.02554(3)$ & $0.045(2)$      \\ 
$0.095$ & $0.10574(3)$ & $0.0776(2)$ & $-0.09(2)$     & $0.515$ & $0.011953(3)$ & $0.02441(3)$ & $0.041(2)$      \\ 
$0.105$ & $0.09752(2)$ & $0.0793(2)$ & $-0.05(2)$     & $0.525$ & $0.011408(3)$ & $0.02327(3)$ & $0.039(2)$      \\ 
$0.115$ & $0.09083(2)$ & $0.0808(2)$ & $-0.007(10)$   & $0.535$ & $0.010884(3)$ & $0.02207(3)$ & $0.038(2)$      \\ 
$0.125$ & $0.08536(2)$ & $0.0816(2)$ & $-0.013(10)$   & $0.545$ & $0.010370(3)$ & $0.02093(2)$ & $0.035(2)$      \\ 
$0.135$ & $0.08073(2)$ & $0.0828(2)$ & $0.021(9)$     & $0.555$ & $0.009869(3)$ & $0.01985(2)$ & $0.032(2)$      \\ 
$0.145$ & $0.07688(2)$ & $0.0832(2)$ & $0.030(9)$     & $0.565$ & $0.009390(3)$ & $0.01875(2)$ & $0.033(2)$      \\ 
$0.155$ & $0.07358(2)$ & $0.0833(2)$ & $0.052(8)$     & $0.575$ & $0.008911(2)$ & $0.01764(2)$ & $0.030(2)$      \\ 
$0.165$ & $0.07082(2)$ & $0.0828(2)$ & $0.042(8)$     & $0.585$ & $0.008449(2)$ & $0.01657(2)$ & $0.028(2)$      \\ 
$0.175$ & $0.06845(2)$ & $0.0817(2)$ & $0.070(8)$     & $0.595$ & $0.007992(2)$ & $0.01554(2)$ & $0.024(2)$      \\ 
$0.185$ & $0.06643(2)$ & $0.0781(2)$ & $0.036(8)$     & $0.605$ & $0.007538(2)$ & $0.01450(2)$ & $0.022(1)$    \\ 
$0.195$ & $0.06320(2)$ & $0.0783(2)$ & $0.049(8)$     & $0.615$ & $0.007089(2)$ & $0.01344(2)$ & $0.021(1)$    \\ 
$0.205$ & $0.05874(2)$ & $0.0808(2)$ & $0.071(7)$     & $0.625$ & $0.006643(2)$ & $0.01244(2)$ & $0.0207(9)$     \\ 
$0.215$ & $0.05478(2)$ & $0.0802(2)$ & $0.079(7)$     & $0.635$ & $0.006192(2)$ & $0.01146(2)$ & $0.0175(8)$     \\ 
$0.225$ & $0.05119(2)$ & $0.0788(2)$ & $0.083(7)$     & $0.645$ & $0.005746(2)$ & $0.01046(2)$ & $0.0153(8)$     \\ 
$0.235$ & $0.04797(1)$ & $0.0768(2)$ & $0.103(6)$   & $0.655$ & $0.005296(2)$ & $0.00948(2)$ & $0.0141(7)$     \\ 
$0.245$ & $0.04505(1)$ & $0.0743(1)$ & $0.099(6)$ & $0.665$ & $0.004830(2)$ & $0.00850(1)$ & $0.0117(7)$   \\ 
$0.255$ & $0.042376(9)$ & $0.07180(9)$ & $0.094(6)$   & $0.675$ & $0.004359(2)$ & $0.00752(1)$ & $0.0105(6)$   \\ 
$0.265$ & $0.039926(9)$ & $0.06926(9)$ & $0.092(6)$   & $0.685$ & $0.003874(2)$ & $0.006593(9)$ & $0.0083(5)$    \\ 
$0.275$ & $0.037697(8)$ & $0.06666(8)$ & $0.097(5)$   & $0.695$ & $0.003369(2)$ & $0.005645(8)$ & $0.0073(5)$    \\ 
$0.285$ & $0.035639(8)$ & $0.06417(8)$ & $0.097(5)$   & $0.705$ & $0.002843(1)$ & $0.004749(7)$ & $0.0061(4)$  \\ 
$0.295$ & $0.033751(7)$ & $0.06167(8)$ & $0.099(5)$   & $0.715$ & $0.0022871(9)$ & $0.003931(6)$ & $0.0056(4)$   \\ 
$0.305$ & $0.031981(7)$ & $0.05915(7)$ & $0.088(5)$   & $0.725$ & $0.0016951(7)$ & $0.003205(5)$ & $0.0046(3)$   \\ 
$0.315$ & $0.030343(7)$ & $0.05690(7)$ & $0.089(5)$   & $0.735$ & $0.0010592(6)$ & $0.002754(4)$ & $0.0044(2)$   \\ 
$0.325$ & $0.028826(7)$ & $0.05464(6)$ & $0.087(4)$   & $0.745$ & $367.5(4) \cd 10^{-6}$ & $0.002889(2)$ & $0.0051(1)$ \\ 
$0.335$ & $0.027407(6)$ & $0.05245(6)$ & $0.077(4)$   & $0.755$ & $0$ & $0.0018054(7)$ & $0.00524(3)$            \\ 
$0.345$ & $0.026087(6)$ & $0.05042(6)$ & $0.092(4)$   & $0.765$ & $0$ & $751.8(3) \cd 10^{-6}$ & $0.00248(2)$            \\ 
$0.355$ & $0.024841(5)$ & $0.04837(6)$ & $0.076(4)$   & $0.775$ & $0$ & $380.1(2) \cd 10^{-6}$ & $0.001189(9)$           \\ 
$0.365$ & $0.023668(5)$ & $0.04642(5)$ & $0.076(4)$   & $0.785$ & $0$ & $191.1(2) \cd 10^{-6}$ & $581(5) \cd 10^{-6}$           \\ 
$0.375$ & $0.022574(5)$ & $0.04445(5)$ & $0.081(4)$   & $0.795$ & $0$ & $89.11(8) \cd 10^{-6}$ & $255(3) \cd 10^{-6}$          \\ 
$0.385$ & $0.021545(5)$ & $0.04283(5)$ & $0.071(3)$   & $0.805$ & $0$ & $35.65(5) \cd 10^{-6}$ & $99(2) \cd 10^{-6}$          \\ 
$0.395$ & $0.020570(5)$ & $0.04100(5)$ & $0.073(3)$   & $0.815$ & $0$ & $10.24(3) \cd 10^{-6}$ & $29.5(8) \cd 10^{-6}$         \\ 
$0.405$ & $0.019638(5)$ & $0.03928(5)$ & $0.067(3)$   & $0.825$ & $0$ & $1.241(8) \cd 10^{-6}$ & $1.6(2) \cd 10^{-6}$        \\ 
$0.415$ & $0.018760(4)$ & $0.03781(4)$ & $0.063(3)$   & $0.835$ & $0$ & $2.3(2) \cd 10^{-9}$ & $139(4) \cd 10^{-9}$     \\
\hline 
\end{tabular} 
}
\end{table}

\clearpage



\end{document}